\documentclass[12pt]{article}

\usepackage{epsfig}
\usepackage{floatflt}

{\end{list}}
\newcounter{enumct}

\newcommand{\newc}{\newcommand}
\newc{\ra}{\rightarrow}
\newc{\pom}  {I\hspace{-0.2em}P}
\newc{\eg}{{\it e.g.}\ }
\newc{\ie}{{\it i.e.}\ }
\newc{\cf}{{\it cf.}\ }
\newc{\lam}{\lambda}
\newc{\eps}{\epsilon}
\newc{\gev}{\,GeV}
\newc{\Ebar}{{\bar E}}
\newc{\Dbar}{{\bar D}}
\newc{\Ubar}{{\bar U}}
\newc{\rp}{$R_p$}
\newc{\half}{\frac{1}{2}}
\newc{\rpv}{{\not \!\! R_p}}
\newc{\rpvm}{{\not  R_p}}
\newc{\eq}[1]{(\ref{eq:#1})}
\newc{\lab}[1]{\label{eq:#1}}
\newc{\gsim}{{\stackrel{>}{\sim}}}
\newc{\lsim}{{\stackrel{<}{\sim}}}
\newc{\nonum}{\nonumber}
\newc{\rpmssm}{{$R_p$-MSSM}\ }
\newc{\rpvmssm}{$\rpv$-MSSM}
\newc{\xgo}{x_\gamma^{\rm OBS}}
\newc{\ETJ}{E^{{\rm Jet}}_T}

\def\3{\ss}

\def\g2{GeV$^{2}$}
\def\ds{D^{\ast}}
\def\d0{D^{0}}

\def\dsk3pi{ {\ds}^{+}~\rightarrow~\d0~\pi^{+}_{S}%
        \rightarrow~(K^{-}~\pi^{+}~\pi^{+}~\pi^{-})~\pi^{+}_{S} }

\def\xgo{x_\gamma^{\rm OBS}}

\begin{document}
 
\sloppy

\pagestyle{empty}

\begin{center}
{\LARGE\bf Charm `in' the Photon}\\[10mm]
{\Large J. M. Butterworth} \\[3mm]
{\it Department of Physics and Astronomy,}\\[1mm]
\psfig{figure=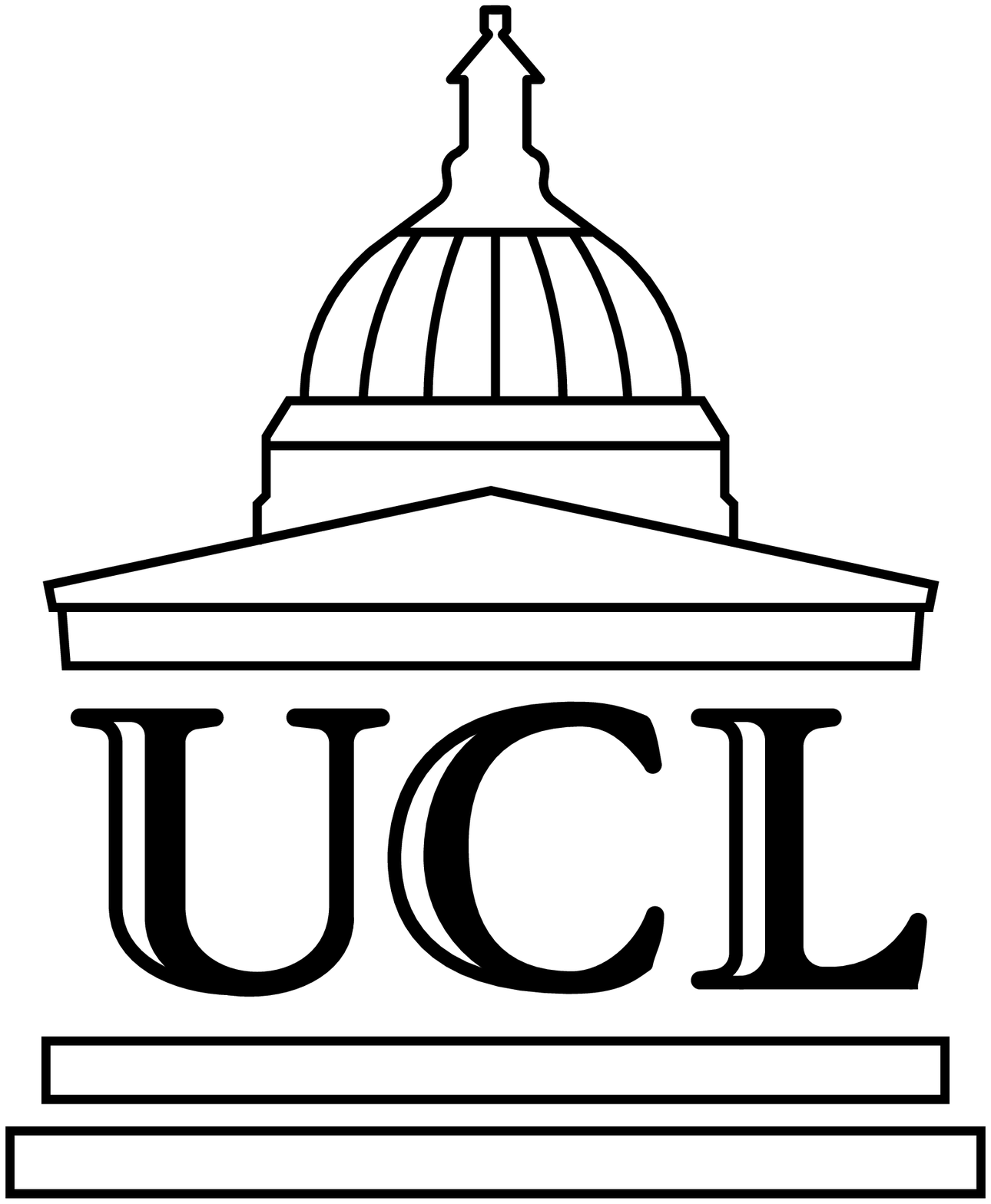,height=0.5cm}\\ 
{\it University College London, UK}\\[1mm]
{\it ZEUS Collaboration}\\[1mm]
{\it E-mail: jmb@hep.ucl.ac.uk}\\[20mm]

{\bf Abstract}\\[1mm]
\begin{minipage}[t]{140mm}
Some recent ZEUS results on the photoproduction of open charm
and possible implications for photon structure are discussed.
\end{minipage}\\[5mm]

\rule{160mm}{0.4mm}

\end{center}

\section{Open Charm Photoproduction at HERA}

Photoproduction of `open' charm (as opposed to $c\bar{c}$ bound
states) can take place via several processes at HERA. The most obvious
is photon-gluon fusion. The photon interacts directly with a gluon
from the proton via $t$-channel charm quark, producing a high
transverse energy charm balanced by an anticharm (Fig.~1a). 

At high enough transverse energies, the $c$ and $\bar{c}$ will each
lead to the formation of jets of hadrons. One hadron in each jet will
in general be charmed.  Photon-gluon fusion is often assumed to be the
dominant, or indeed only, process.

Nevertheless, other possible production mechanisms exist. The photon
can fluctuate into a $q\bar{q}$ state which may be long lived on the
timescale of strong interactions. The $q\bar{q}$ state can thus form a
complex partonic structure. Partons from the photon can then undergo
hard scattering with partons from the proton - so called `Resolved
Photon' interactions. This allows the possibility of charm production
via gluon-gluon fusion (Fig.~1b), where one gluon comes from the
proton, the other from the photon.

\begin{figure}[ht]
\psfig{file=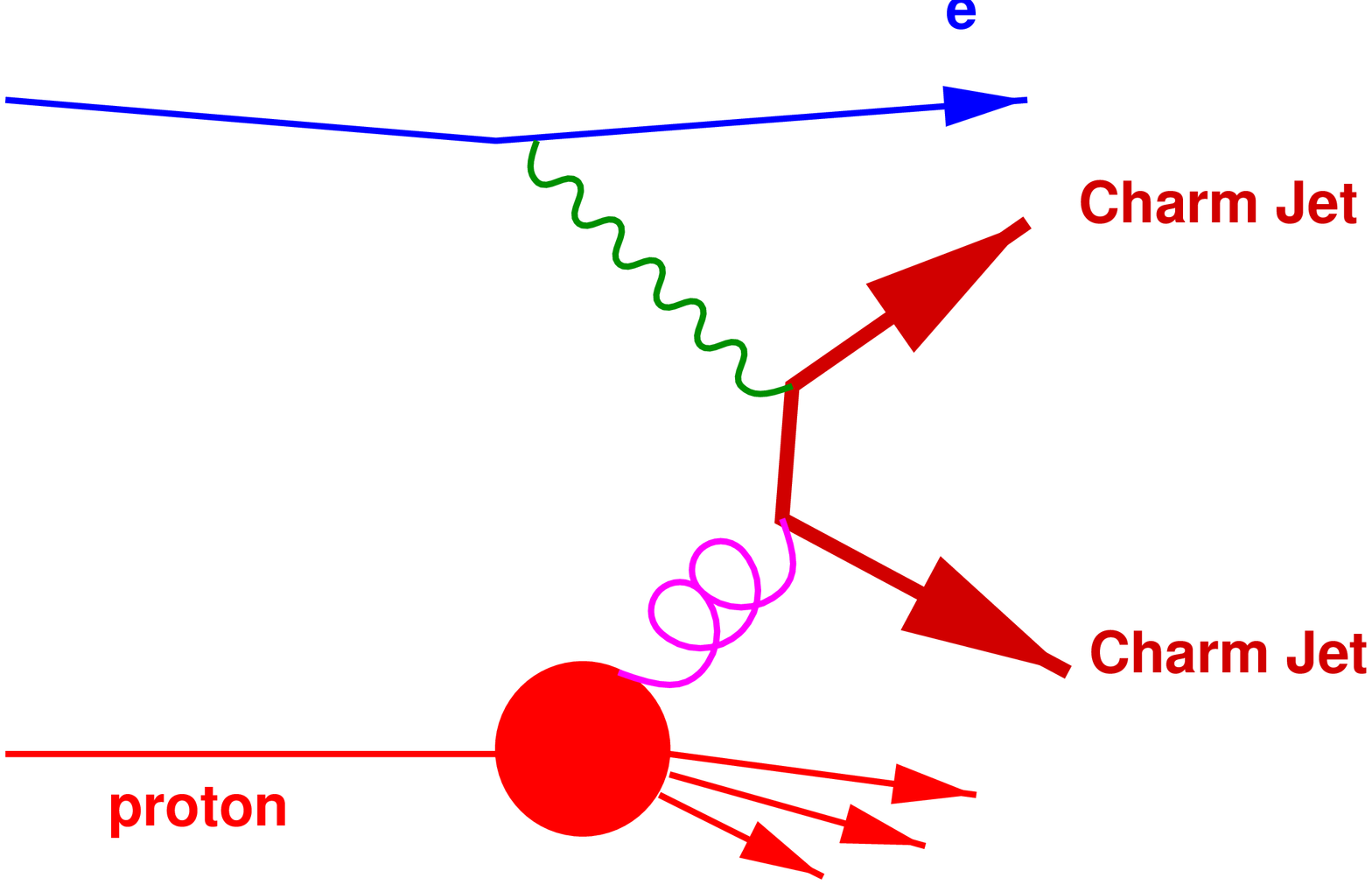,height=5.0cm,angle=270}\psfig{file=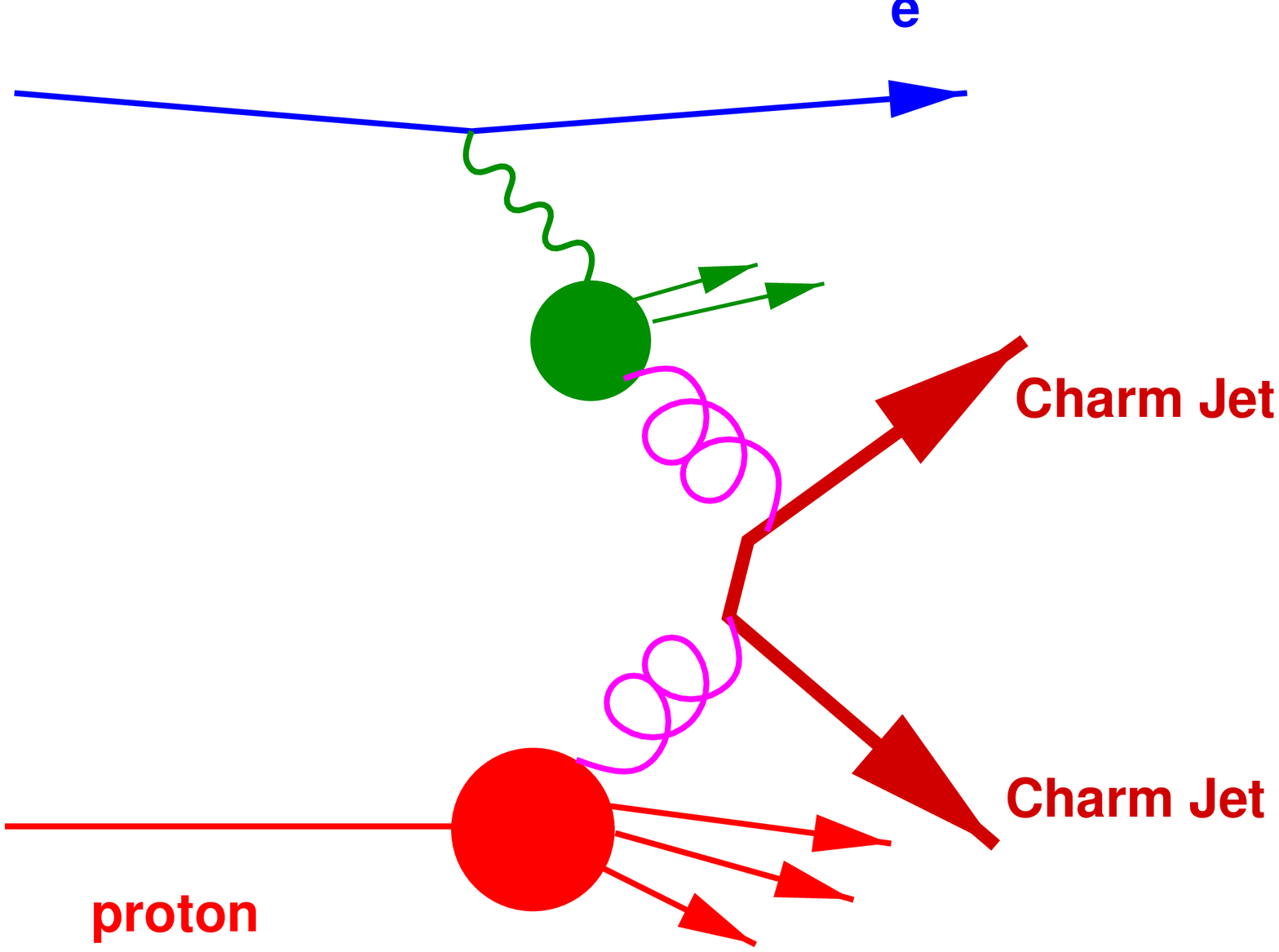,height=5.0cm,angle=270}\psfig{file=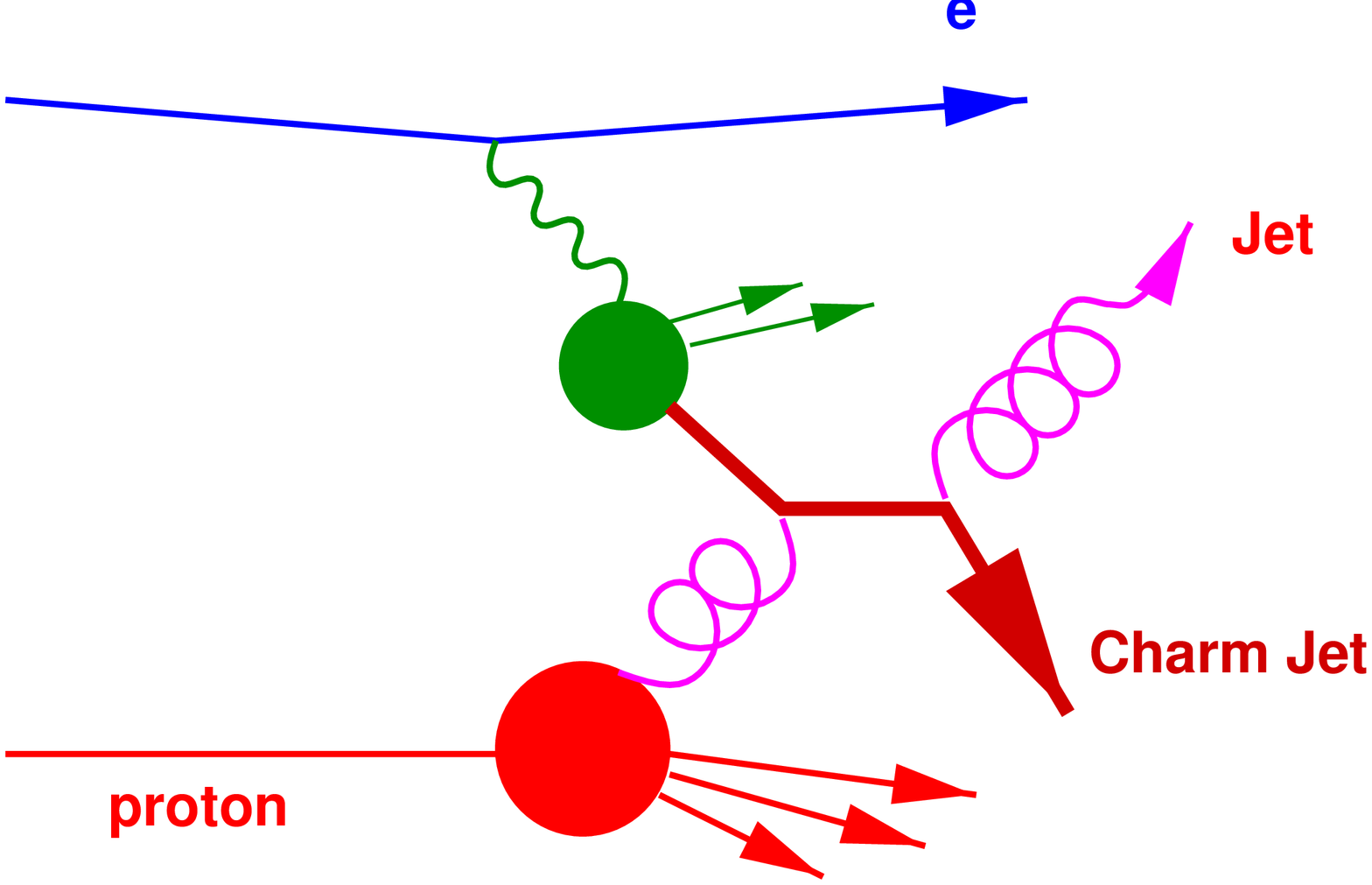,height=5.0cm,angle=270}
\caption{\it a) Photon Gluon fusion b) Gluon Gluon fusion c) Charm excitation.}
\end{figure}

This process is very similar to direct photoproduction, producing two
high $\ETJ$ jets each containing a charmed hadron. The difference is
that only a fraction of the photon's momentum enters into the jet
production, the rest being carried off in a photon remnant.

A further class of resolved processes can be imagined. What if the
parton structure evolved by the photon contains charm? In this case,
so-called `charm excitation' processes can take place (Fig.~1c).

This process also leads to two high $\ETJ$ jets, but only one of them
contains a charmed hadron. The second charmed hadron is carried off in
the photon remnant.

It is interesting to ask whether the resolved diagrams are
important. The contribution of the first is especially sensitive to
the gluon distribution inside the photon, whereas the second addresses
the question as to whether charm is somehow generated `inside' the
photon.

We should ask how charm could be generated `in' the photon. Might it
happen via $\gamma \rightarrow c\bar{c}$ or $g \rightarrow c\bar{c}$?
Is it perturbatively calculable?  One has to be careful to define what
exactly is meant by charm inside the photon. It is important to note
that at NLO the division between `charm excitation' and `boson gluon
fusion' (and indeed between resolved and direct photoproduction in
general) will depend upon the choice of factorization scale.  Moving
factorization scale can turn a LO charm excitation diagram into a NLO
direct photoproduction diagram, as illustrated in Fig.~2. A similar
arbitrariness is present between charm generated via gluon splitting
or assigned to the gluon-gluon fusion process.

\begin{floatingfigure}[l]{13.1cm}
\psfig{file=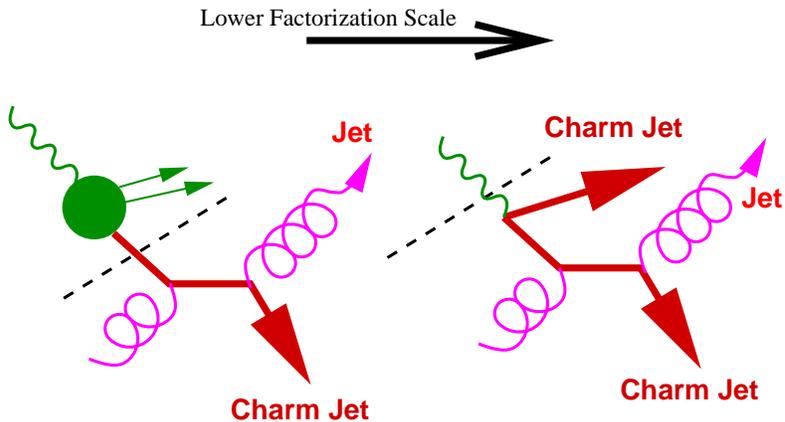,width=5.5cm,angle=270}
\caption{\it NLO confusion}
\end{floatingfigure}

Thus the discussion of any results will to some extent depend upon
what approximations are used in the calculations to which the data is
being compared.

\subsection{Massless or Massive?}

Currently two different approximations are used in the calculation
of charm photoproduction at next-to-leading order.

\begin{itemize}

\item{\bf Resummed, or `Massless'.}  This approach uses ${\cal
O}(\alpha_s^2)$ matrix elements for charm treated as a massless parton
over the threshold for its production.  Charm is an active flavour in
the photon, present at a level dependent upon the choice of parton
distribution set. By allowing charm to be generated in the evolution
of the photon parton distribution, this approach resums logarithms of
$\ETJ /m_c$~\cite{cacc,kniehl}.

It is expected that this scheme will be a good approximation at $\ETJ
\gg m_c$.

\item{\bf Massive.} In this approach, ${\cal O}(\alpha_s^2)$ matrix
elements for massive charm are used. There is no charm content
assigned to the parton distributions inside the photon. No resummation
of $\ETJ /m_c$ logarithms is performed~\cite{frix}.

It is expected that this scheme will be a good approximation when 
$\ETJ \approx m_c$.

\end{itemize}

In the jet measurements to be discussed below~\cite{zeusdstar}, $\ETJ
\approx 7$~GeV.

\subsection{Measuring Charm}

The most commonly used method so far for tagging charm at HERA is the
$D^*$ tagging method~\cite{dstarmeth}. This technique exploits the fact
that the mass difference between the $D^*$ and $D^0$ is small. Thus by
cutting on this reconstructed mass difference as well as the mass, a
relatively pure sample of charmed events is obtained.

\section{Inclusive $D^*$ Cross Section}

A sample of charm events is selected using the following cuts;

\begin{floatingfigure}{8cm}
\psfig{file=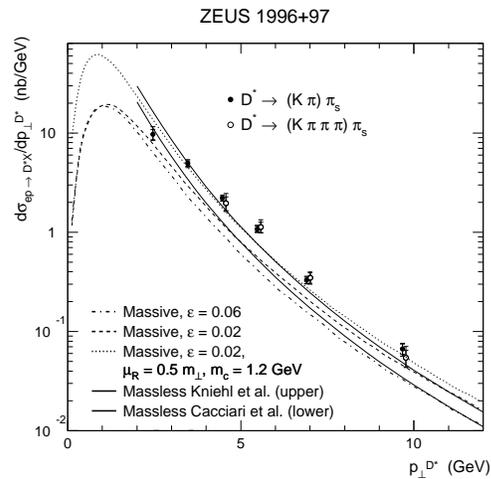,width=6.5cm}
\caption{\it $d\sigma/dp_T(D^*)$
The MRSG~\cite{MRSG} and GRV-G~HO~\cite{GRV} parton density functions
are used for the proton and the photon respectively.}
\end{floatingfigure}

$\bullet ~p_T(D^*) > 2.0$~GeV, (when the $D^0$ decays to $K\pi$) or
 $p_T(D^*) > 4.0$~GeV, (when the $D^0$ decays to $K\pi\pi\pi$)

$\bullet ~ |\eta (D^*)| < 1.5$.

$\bullet ~ 130 < W_{\gamma p} < 280$~GeV

$\bullet ~ Q^2 < 1$~GeV$^2$

The differential cross section $d\sigma/dp_T(D^*)$ is shown in Fig.~3,
compared to various NLO QCD calculations.

It can be seen that even with an extreme choice of parameters such as
the charm mass, the massive scheme tends to lie below the data (dotted
line). In addition, there is a significant discrepancy between the two
groups using massless charm calculations.

The differential cross section $d\sigma/d\eta(D^*)$ is shown in Fig.~4.

\begin{figure}
\begin{center}
\psfig{file=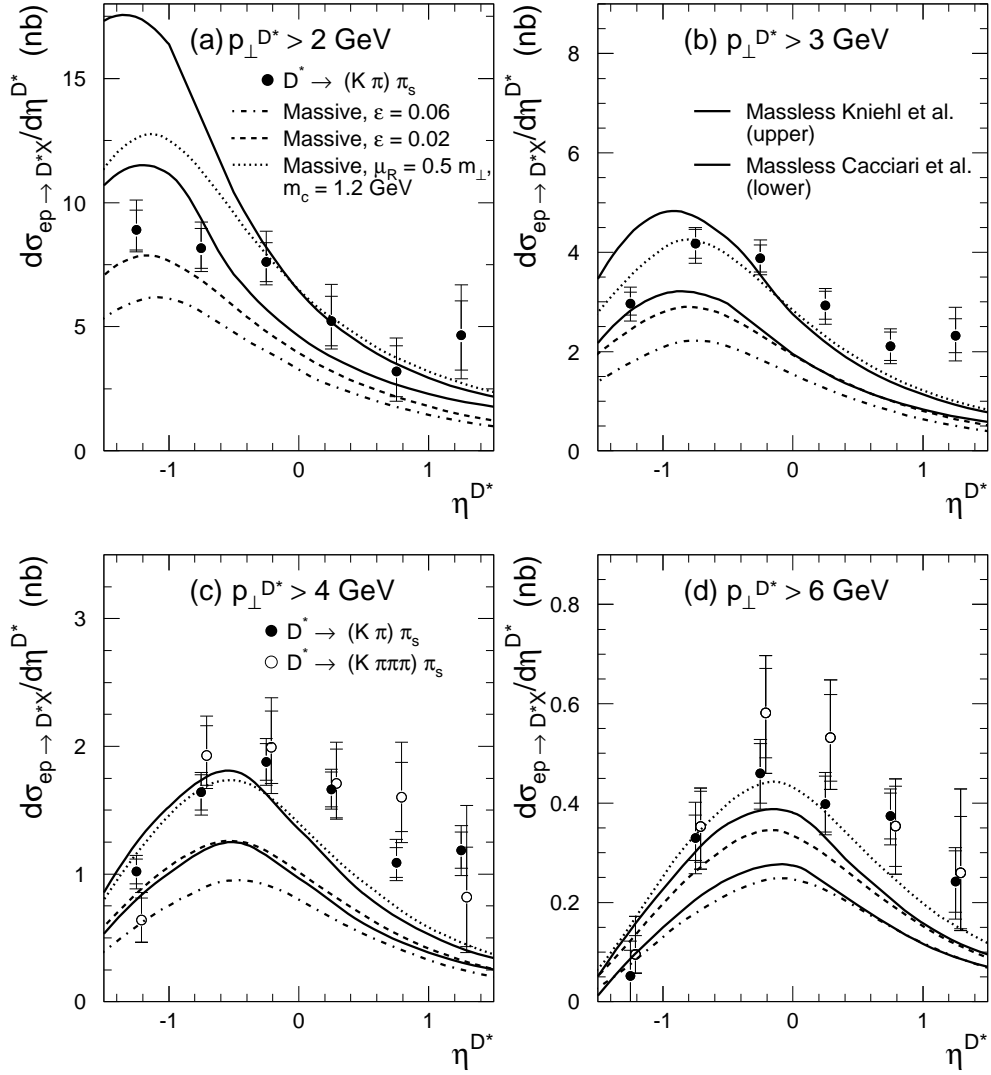,width=13cm}
\end{center}
\caption{\it $d\sigma/d\eta(D^*)$
The MRSG~\cite{MRSG} and GRV-G~HO~\cite{GRV} parton density functions
are used for the proton and the photon respectively.}
\end{figure}

Again, the massive calculations generally lie below the data,
especially in forward direction, and the discrepancy between the two
different massless calculations is clear. In the massless scheme, this
cross section has a sensitivity to the parton distributions in the
photon which is of the same order as the other uncertainties at
present.

\section{Photoproduction of Charm in Jets}

Further information about the charm production mechanism can be
obtained by measuring jets in high $\ETJ$ photoproduction and looking
for charm inside the jets. This has been done, again using the $D^*$
tagging method, in a similar $W$ range and with $\ETJ > 6$~GeV,
$p_T(D^*) > 3$~GeV.  The jets are defined using the $K_T$
algorithm~\cite{catani} in `inclusive' mode~\cite{ellis}.

Especially given the excess of data over the massive charm
calculations, it is of course interesting to separate direct and
resolved samples. This is possible using the variable~\cite{xgo}:
\[
\xgo = \frac{\Sigma_{\rm jets}(\ETJ\ e^{-\eta^{jet}})}{2 E_{e} y}
\]
which is the fraction of the photon's momentum which participates in
the production of the two highest $\ETJ$ jets. Thus LO direct process
have $\xgo =1$ and LO resolved processes have lower $\xgo$, although
$\xgo$ itself is of course defined independently of the order of the
calculation.

Fig.~5 shows the energy flow around jets, compared to the expectation
from the HERWIG Monte Carlo.  The energy flow in the rear (negative
$\Delta\eta$) region shows evidence for presence of a photon remnant
in a significant fraction of the events at low $\xgo$.

\begin{floatingfigure}{12cm}
\begin{center}
\psfig{file=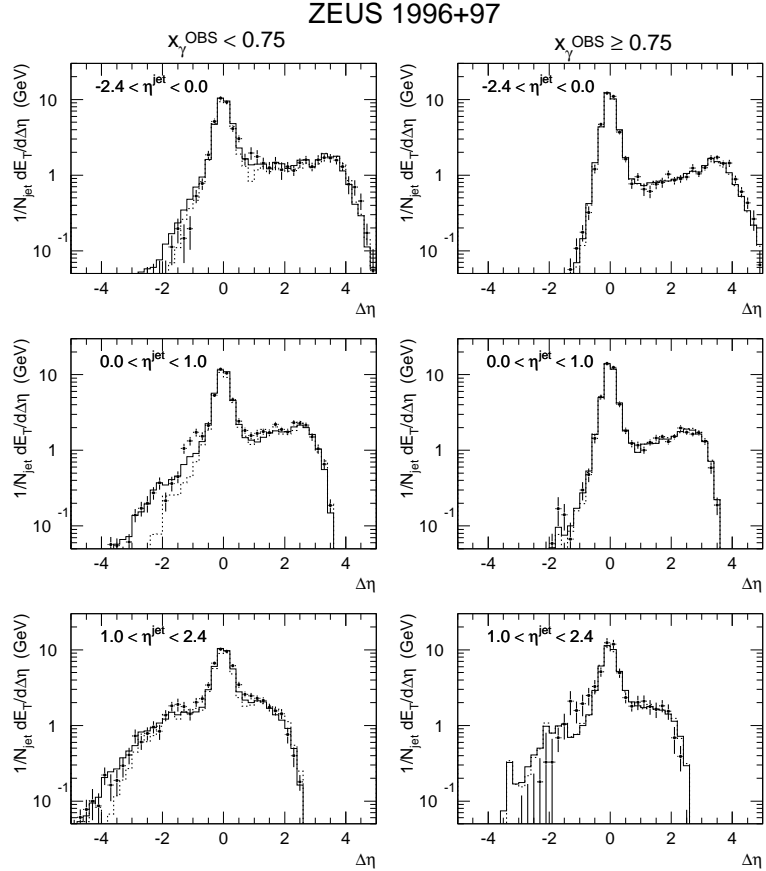,width=10cm}
\caption{\it Energy flow around jets}
\end{center}
\end{floatingfigure}

The cross section $d\sigma/d\xgo$ is shown in Fig.~6. The cross
section at lower $\xgo$ values is significant, indicating that LO
direct processes alone cannot describe charm production successfully.
In particular, the data is inconsistent with the LO Direct process in
HERWIG shown in the figure, even after the effects of parton showering
and hadronisation are included in the Monte Carlo. Such effects can
populate the low $\xgo$ region even with direct events, but do not do
so sufficiently. In fact the data require a ($45 \pm 5$)\% LO resolved
contribution from HERWIG.

Furthermore, according to HERWIG this resolved contribution is almost
entirely charm excitation.

Also shown, in the lower half of Fig.~6, is a NLO massive calculation
of $d\sigma/d\xgo$. The calculation lies below the data at low
$\xgo$. It should be remembered that hadronisation is not included in
the NLO calculation and this may affect the comparison. Nevertheless,
the data suggest a larger resolved contribution than is present in the
calculation.

\section{Summary}

A significant cross section for the `resolved' photoproduction of
charm has been measured. The theory is `close but no cigar' in
inclusive $D^*$ and charmed jet measurements, lying in general
somewhat below the data particularly in the forward and low-$\xgo$
regions.

Comparison to the HERWIG simulation, which includes LO matrix
elements, leading logarithmic parton shower and a hadronisation
model, requires a charm excitation contribution of about 45\% in the
kinematic regime measured here.

The parton distribution functions used in the theory comparisons do
not always represent the state-of-the-art in their massive quark
treatment, and our understanding should benefit from a comparison to
other parameterisations.

The data represent a challenge to the theory to truly understand charm
production `inside' the photon. This challenge is likely to get
tougher as more accurate measurements over wider kinematic regimes
become possible from both H1 and ZEUS with micro-vertex detectors, the
introduction of other tagging methods, and higher luminosity from the
coming HERA upgrade. I would like to acknowledge the enormous
efforts of the ZEUS heavy flavour group, as well as the theory groups
who provided the calculations, many of whom will undoubtedly be
responsible for these coming advances.

\begin{figure}[ht]
\begin{center}
\psfig{file=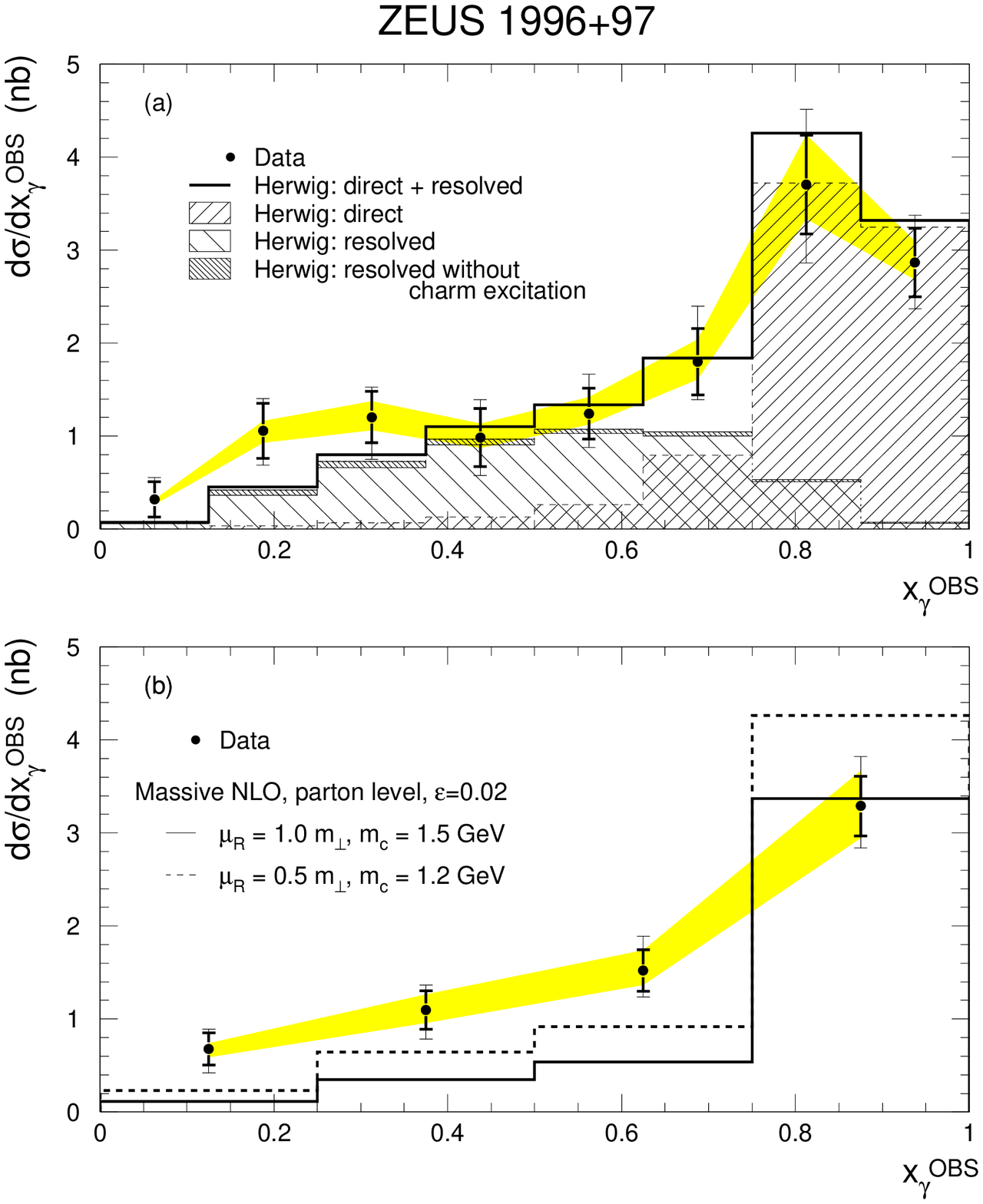,height=18cm}
\end{center}
\caption{\it $d\sigma/d\xgo$
The MRSG~\cite{MRSG} and GRV-G~HO~\cite{GRV} parton density functions
are used for the proton and the photon respectively.}
\end{figure}

\end{document}